# Single-Atom Scale Structural Selectivity in Te Nanowires Encapsulated inside Ultra-Narrow, Single-Walled Carbon Nanotubes

Paulo V. C. Medeiros,*,† Samuel Marks,‡ Jamie M. Wynn,† Andrij Vasylenko,‡,¶ Quentin M. Ramasse,§ David Quigley,‡,∥ Jeremy Sloan,*,‡ and Andrew J. Morris*,†,‡

†*Theory of Condensed Matter Group, Cavendish Laboratory, University of Cambridge, J. J. Thomson Avenue, Cambridge CB3 0HE, U.K.*
‡*Department of Physics, University of Warwick, Coventry CV4 7AL, U.K.*
¶*Institute for Condensed Matter Physics, National Academy of Science of Ukraine (NAS Ukraine), 1 Sventsitskii street, 79011 Lviv, Ukraine*
§*SuperSTEM Laboratory, STFC Daresbury, Keckwick Lane, Daresbury WA4 4AD, U.K.*
∥*Centre for Scientific Computing, University of Warwick, Coventry CV4 7AL, U.K.*

E-mail: pvm20@cam.ac.uk; j.sloan@warwick.ac.uk; ajm255@cam.ac.uk

**Abstract**

Extreme nanowires (ENs) represent the ultimate class of crystals: They are the smallest possible periodic materials. With atom-wide motifs repeated in one dimension (1D), they offer a privileged perspective into the Physics and Chemistry of low-dimensional systems. Single-walled carbon nanotubes (SWCNTs) provide ideal environments for the creation of such materials. Here we present a comprehensive study of Te ENs encapsulated inside ultra-narrow SWCNTs with diameters between $0.7\,nm$ and $1.1\,nm$. We combine state-of-the-art imaging techniques and 1D-adapted *ab initio* structure prediction to treat both confinement and periodicity effects. The studied Te ENs adopt a variety of structures, exhibiting a true 1D realisation of a Peierls structural distortion and transition from metallic to insulating behaviour as a function of encapsulating diameter. We analyse the mechanical stability of the encapsulated ENs and show that nanoconfinement is not only a useful means to produce ENs, but may actually be necessary, in some cases, to prevent them from disintegrating. The ability to control functional properties of these ENs with confinement has numerous applications in future device technologies, and we anticipate that our study will set the basic paradigm to be adopted in the characterisation and understanding of such systems.

Extreme nanowires (ENs)[1] are few-atom-wide wires that extend along one dimension and have their fine structural details resolved also at the atomic length scale. They represent the smallest possible one-dimensional (1D) materials – a fact which, on its own, justifies detailed experimental and theoretical investigation into materials structure and characterisation. In contrast to exotic materials, predicted *via* computational discovery, practical routes to synthesis have revealed a rich landscape of one-dimensional Physics and Chemistry. Such extreme 1D structures have an obvious potential for use as interconnecting elements between components in nanocircuits, but less immediate applications, such as the design of nano phase-change materials (NPCMs) for use in high-efficiency solid-state storage devices, can be envisioned, especially since structural changes in 1D can be induced – and precisely characterised, as we show here – by the equally extreme confinement of matter at the nanoscale.

A practical route to achieving nano confinement in 1D is making use of single-walled carbon nanotubes (SWCNTs). A prominent characteristic of SWCNTs is their capillarity,[2] which has been explored in experiments to grow and encapsulate crystals, molecules and aggregates. SWCNTs have remarkable chemical stability and provide a very effective chemical shielding from the external environment. These characteristics make such tubes the ideal templates for the discovery of nanomaterials, as well as for the investigation of new material phases that emerge due to nano 1D confinement.[3–7] Conversely, we have recently shown that encapsulated ENs can be used to boost electronic transport in SWCNTs.[8] This further highlights the importance of such materials for applications in electronics, specially with the demonstrated possibility of building SWCNT-based transistors with gate lengths as small as 1 $nm$.[9] Hybrid nanotube-nanowire systems have been used in the fabrication of nanothermometers[10] and magnetic force microscopy sensors,[11] and potential for many other applications,[12] such as battery electrodes[13] and in photothermal nanomaterials-based devices,[14] has also been demonstrated.

Applications of alloys containing Ge, Sb and Te in phase-change materials have enjoyed considerable success over the past decades,[15] and the search for NPCMs based on such elements is a natural step.[16] Te is of particular interest in this context because its bulk structure favours the formation of 1D structures.[17,18] Ultranarrow Te nanowires down to 25 $nm$ in diameter exhibit a wealth of interesting properties related to their photoconductivity, nonlinear optical response, thermoelectric, and piezoelectric effects.[19,20] They have been employed in gas sensors,[21] optoelectronic devices,[22] and photonic crystals.[23] Recently, the dimensionality of Te wires was reduced to single atom chains in carbon nanotubes (CNTs) with *internal* diameters in the range 1.2 ± 0.2 $nm$.[24] There is, however, a clear motivation to produce a more consistent, atomically regulated material, ideally encapsulated exclusively within single conformation ultranarrow single-walled carbon nanotubes (UNSWCNTs) with diameters smaller than 1.2 $nm$. Such UNSWCNTs would have precisely tailored physical properties, being either pure semimetallic or pure semiconducting confining tubes with a single discrete gap.

Recent major developments in hardware for Aberration Corrected High-Resolution Transmission Electron Microscopy (AC-TEM) and Aberration Corrected Scanning Transmission Electron Microscopy (AC-STEM)[25,26] have driven considerable interest in low dimensional materials, as imaging such systems requires very high precision. Such advancements, on the other hand, mean that three-dimensional data with atomic dimension accuracy is now needed for ENs if one is to ensure maximum compatibility between theory and experiment. From the theory standpoint, such developments are paralleled by the advent of high-performance computing, which, allied with clever approaches to structure searching, can increase the efficiency and accuracy in materials discovery and eliminate guesswork from the process.

Here, we combine AC-TEM, AC-STEM and electron energy loss spectroscopy (EELS) measurements with a high-throughput, 1D-adapted implementation of the *ab initio* random structure searching (AIRSS) method[27] to provide the most accurate characterisation to date of Te ENs grown inside UNSWCNTs. The AIRSS methodology has successfully been employed to tackle problems as diverse as the structural determination of point-defects,[28–30] prediction of high-pressure phases,[31] as well as to detect structural transformations in nano-structured silicon-based lithium ion batteries.[32]

## Results and Discussion

We have filled UNSWCNTs with diameters spanning 0.7 $nm$ to 1.1 $nm$ with Te nanowires using a similar sublimation method to that applied by Senga et al.[33] to the filling of double-walled carbon nanotubes with ionic CsI single atom chains. Elemental Te was filled by sublimation *in vacuo* (see Experimental Methods section) directly into the UNSWCNTs to produce single atom thick Te ENs. Our AC-TEM imaging results (Fig. 1) show that the vast majority of the obtained Te ENs consisted of single-atom thick Te 3-fold-symmetry helical coils (3Hs). We found that the "pitch" of each coil varies in a non-monotonic fashion with respect to the observed diameter of the encapsulating UNSWCNT. See the included Supporting Information (SI) for more details. Scanning TEM (STEM) imaging and EELS (Fig. 2(a)-(d)) confirmed the chemical identity of the obtained Te coils as elemental Te.

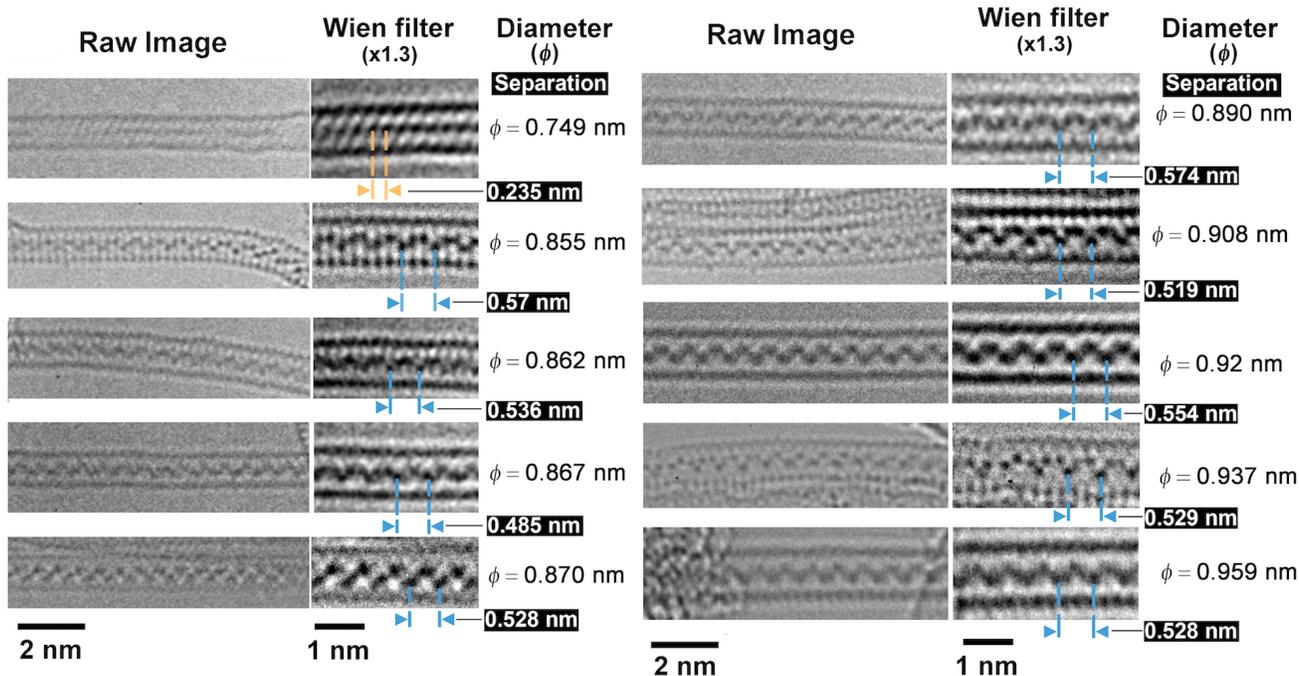

Figure 1: AC-TEM images of UNSWCNTs, sorted by diameter (smallest to largest), filled with 1D Te extreme nanowires (ENs). With the exception of the linear chain encapsulated inside the smallest diameter (0.749 nm) UNSWCNT (Te-Te spacing indicated in orange), all of the remaining ENs form coil structures with differing periods (coil pitches) along the lengths of each nanotube. UNSWCNT diameters and Te atomic period measurements were obtained from the indicated Wien filtered sections with an estimated precision of ±0.008 nm.

In order to unambiguously determine the structures of the encapsulated Te ENs, we adapted the AIRSS method[27] to the modelling of SWCNT-encapsulated 1D structures (see SI). Additionally, we note that the determination of the most favourable ENs structure encapsulated inside a single SWCNT is not enough to provide a precise characterisation of SWCNT-encapsulated ENs.

The formation of distinct EN structures can be favoured by modifying the diameter of the encapsulating SWCNT, and the ability to predict the most favourable structures inside SWCNTs of any given diameter is thus paramount. Particularly at the very small length scale characteristic of confinement inside UNSWCNTs, drastic changes in the shapes of the encapsulated ENs are expected to occur at certain critical UNSWCNT diameters, and a map of such structural changes can be drawn by following the best diameter-dependent encapsulated structures. We have thus selected the most energetically favourable structures obtained from our AIRSS screening on Te ENs and produced the diagram shown in Fig. 3. As discussed in the SI, to deal with strain and mismatch between the ENs and the encapsulating UNSWCNT, the computational procedure to obtain such a diagram involves the use of implicit SWCNTs (ICNTs). We discuss the ICNT model in the SI.

We can assert by analysing the diagram in Fig. 3 that the only viable geometry for the Te EN encapsulated inside UNSWCNT with diameters below 0.77 nm is that of a linear chain (LC). The threshold of 0.77 nm for the emergence of the LC structure is compatible with the van der Waals (vdW) radii of C and Te, which are approximately $0.17\,nm$ and $0.21\,nm$, respectively. Repulsion between the electronic clouds of the encapsulated Te atoms and the UNSWCNTs walls increases rapidly for diameters below the calculated threshold, and the LC configuration is the one that minimises such an overlap. Due to interactions between electron beam and the imaged specimins, we could not ascertain the existence of a Peierls distortion (PD)[35] directly from our experimental data. In our simulations, however, we establish that such distortion is required for the encapsulated Te LC to be mechanically stable – and, therefore, observable (see Fig. 4). The calculated bond length alternation (BLA) is of approximately $0.02\ nm$, leading to a reduction of about $4\,meV$/Te in the total energy of the chain. This BLA does not vary significantly with the diameter of the encapsulating SWCNT. The characteristic effect of a PD in the electronic structure of a LC is the lifting of band degeneracies at the boundaries of the Brillouin zone (of the new 2-atom primitive unit cell (PC)). This is also illustrated in Fig. 4 for the present case. Note, however, that the Te LC remains metallic despite the presence of the PD, as one of the electronic bands continues to cross the Fermi level. Given that PDs are almost exclusively associated with metallic-semimetallic electronic transitions in the literature, the metallic behaviour we predict for the encapsulated Te LC might seem contradictory at first. We point out, nonetheless, that atomic chains exhibiting metallic behaviour albeit featuring PDs have previously been predicted in different contexts.[36,37] We have also found that the Te LC cannot form in vacuum, as it has an unstable transverse acoustic vibrational mode. Such an instability, however, is quenched upon encapsulation, making the encapsulated chain stable. The calculated phonon spectra for the chain in both cases is shown in Fig. 5a. The Te LC is therefore an example of an EN for which nanoconfinement is not only a sufficient condition for mechanical stability, but also a necessary one.

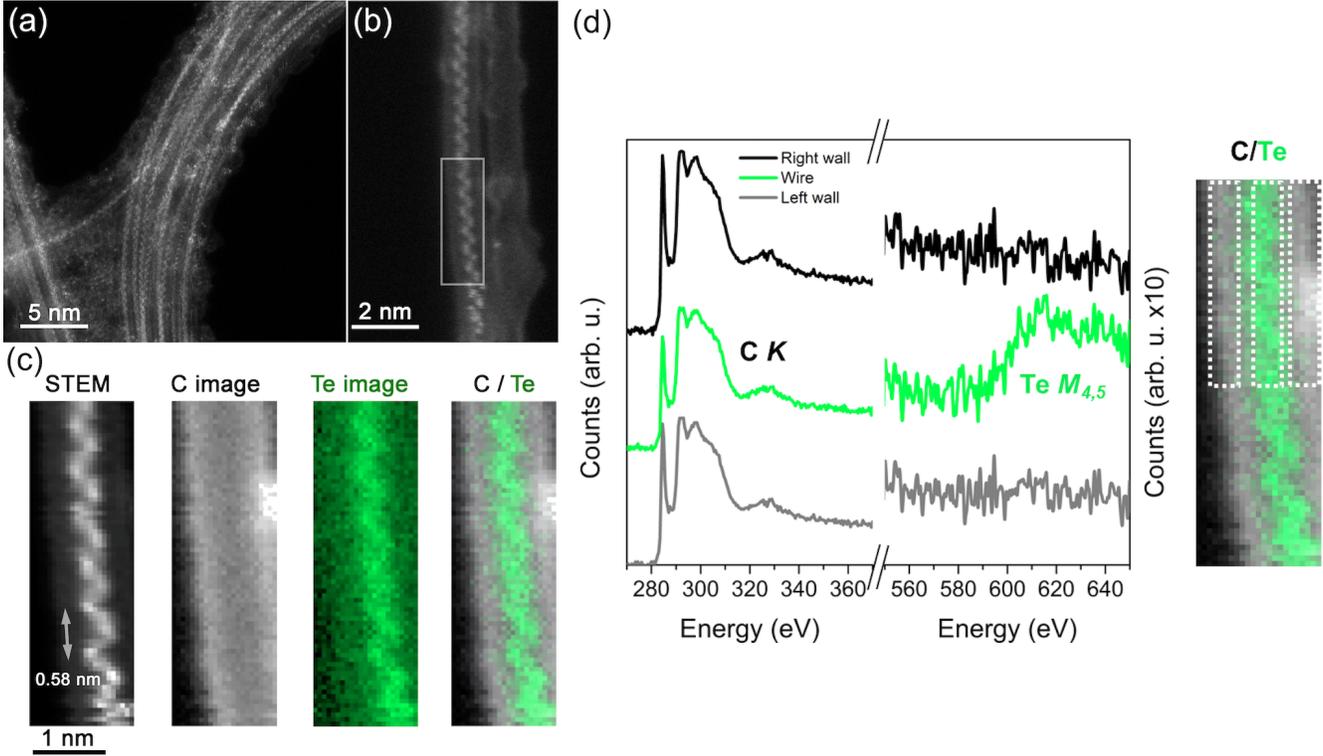

Figure 2: STEM and EELS. (a) Medium magnification STEM ADF image of narrow bundles of UNSWCNTs filled with Te 3H ENs. Filling percentage of UNSWCNTs was determined to be ca. 70%. (b) Higher magnification STEM ADF image showing two UNSWCNTs, one filled with a Te coil (left) and one empty. The boxed region was used for the spectrum imaging recorded in (c) and the EELS spectrum recorded in (d). (c) STEM image, C, Te and combined C/Te spectrum images recorded from the boxed region in (b), confirming the chemical identity of the Te coil. (d) C K edge EELS spectrum and Te M 4,5 edge EELS spectra recorded from the regions indicated in the corresponding detail (inset, right). We found no indication in our EELS measurements that oxygen is present in the immediate vicinity of the wires.

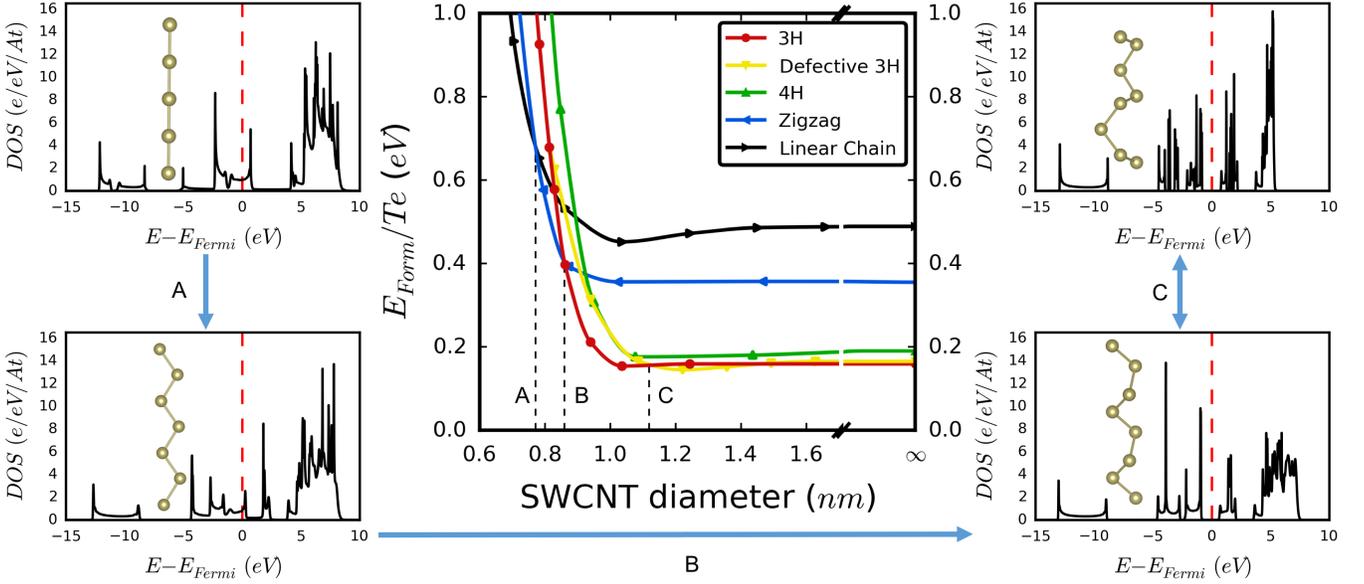

Figure 3: Competing structures of SWCNT-encapsulated Te extreme nanowires. Central panel: Formation energy per encapsulated atom as a function of the encapsulating diameter. The labels A-C mark the diameters at which we predict structural transitions to occur. Left and right panels: Geometries of the structures involved in the transitions A-C, along with the corresponding densities of states (DOS). The competing structures of the ENs were determined using the AIRSS method. The DOS were calculated using the `OptaDOS` code.[34]

We predict the first Te EN structural transition to occur in the vicinity of $d = 0.77\,nm$ (labelled A in Fig. 3), shortly followed by another one at $d \approx 0.86\,nm$ (marked B). From A, our calculations indicate that the zigzag (ZZ) chain configuration becomes accessible for the encapsulated Te ENs, remaining so until the diameters of the encapsulating UNSWCNT reach approximately $0.86\,nm$. With $n$ and $m$ defined as discussed in the SI, we can narrow the possible UNSWCNTs possessing diameters in the A-B interval down to (8, 3), (10, 0), (9, 2), (6, 6), (7, 5), (10, 1), (8, 4), and (9, 3), in order of increasing diameters. The choice $m \geq 0$ (right-hand chirality SWCNTs) was made without loss of generality as we verified that the chirality of the encapsulating UNSWCNTs does not affect the energetics of the encapsulated systems. As in the case of the Te LC, we found that the Te ZZ chain is mechanically unstable in vacuum and therefore cannot form in such an environment. Once more, however, the chain is stabilized upon encapsulation, as shown in Fig. 5b. Such chains therefore provide another example of ENs that require nanoconfinement to achieve mechanical stability.

From B, we predict the ZZ configuration to become less likely to occur, and the 3H form of Te[38] – the 1D building block of the bulk phase of the element – becomes the dominant structure. Depending on whether this 3H building block has right- or left-hand screw symmetry, bulk Te belongs to space groups 152 or 154, respectively. We have found that both chiral orientations of the 3H structure have practically identical formation energies, and are thus equally likely to form inside UNSWCNTs. Interestingly, we found that a defective version of the 3H structure becomes energetically competitive for SWCNT diameters starting from ~ $1.1\,nm$ (marked C in Fig. 3). Such a structure can be seen as a combination of two intercalated ideal 3H chains – one possessing left- and another right-handed screw symmetry. Both the ideal and defective 3H chains are mechanically stable, as can be seen from their phonon spectra presented in Fig. 5d. The fact that defective forms of the 3H EN are energetically favourable compared with the ideal configuration indicates the possibility of triggering structural changes, through the use of external stimuli, to displace Te atoms from their ideal positions. This could be performed, for instance, by irradiation with high-energy electron beams – although one has to exercise care in doing this, as we found in our experiments that the Te chains can become highly mobile, or even unstable, when interacting with high-energy electron beam radiation. Such a high mobility even caused the encapsulating SWCNTs to become empty again in a few occasions. Finally, we point out that, given the proximity of the curves representing the 3H and ZZ structures in Fig. 3, such structures can reasonably be expected to coexist at the range of diameters between A and B.

In the left- and right-hand side panels in Fig. 3 we present the electronic densities of states for the Te structures involved in the transitions A-C. Owing to the general chemical inertness of SWCNTs, and to the particular tendency shown by Te to form covalent intra-chain bonds and only weak, vdW-mediated inter-chain connections, the effects of EN-SWCNT interactions on the electronic structure of the encapsulated ENs are small. By inspecting the plots, we conclude that the structural change marked B also represents a transition from a metallic to a semiconducting EN state. Since encapsulating ENs inside SWCNTs is analogous to submitting the ENs to external pressure, the possibility demonstrated here of inducing metal-semiconductor transitions by encapsulation of Te ENs inside UNSWCNTs means that encapsulation can provide a controllable route towards the construction of pressure-stabilised on-off switches based on Te ENs – despite the semiconducting nature of bulk Te.

Apart from the ground state structures discussed so far, our AIRSS screening also yielded a number of metastable structures. From such structures, we selected those whose curves in the structure diagram (Fig. 3) are located sufficiently close, at least at one point, to the curves corresponding to the diameter-dependent ground state geometries. We consider energies to be sufficiently close when they are separated by about $1\,k_BT$, where $k_B$ is the Boltzmann constant and $T$ represents the standard room temperature. At such a temperature, $k_BT \approx 25.7\,meV$. One of such metastable structures is a double LC (two parallel LCs), which we found to be energetically accessible at diameters around $0.85\,nm$. Upon further inspection, however, we found that the double LC has a mechanically unstable longitudinal acoustic vibrational mode that cannot be stabilised by encapsulation. For this reason, we have not included this structure in our diagram. We verified that such a mode does not represent a Peierls instability. The other metastable structure is a 4-fold-symmetry helix (4H) akin to the one previously suggested for encapsulated sulphur.[39] Although we have not observed such structure in our experiments, we have determined that it is mechanically stable upon encapsulation (see Fig. 5c), and could thus be accessible in small fragments. Our structure diagram indicates that this could occur for encapsulating SWCNTs diameters around $1.1\,nm$ – the upper limit for the SWCNT diameters we considered experimentally.

Fig. 6 presents a comparison between our experimental and simulated TEM images for the encapsulated Te 3H EN. The model in the simulated image corresponds to the most energetically favourable structure we predict for an encapsulating SWCNT with a diameter of $0.949\,nm$, while the experimental TEM image was obtained for encapsulated Te inside a SWCNT with an estimated diameter of $0.945 \pm 0.015\,nm$. Images (a.I) and (a.III) clearly represent domains inside the encapsulating SWCNT in which the structure shown in (b) has been formed. Most interestingly, however, is the observation that, although the structure in image (a.II) looks distinct from (a.I) and (a.III), and could thus, in principle, correspond to a different EN structure, it matches exactly the simulated image (c), which has been obtained using the same model as in (b), differing only by a rotation around the translation axis. We can therefore affirm that (a.II) also corresponds to an encapsulated Te 3H EN. The very low frequencies (Fig. 5d) of both the ideal and defective 3H acoustic torsional vibration modes (modes that correspond to a rigid rotation of the entire system around its symmetry axis at the limit $\boldsymbol{q} \to \boldsymbol{0}$) indicate that the energy cost to twist such structures around their symmetry axes is very small (for reference, $1000\,cm^{-1}$ corresponds to approximately $124\,meV$). Indeed, the different structural domains found in our experiments for the 3H chain, as depicted in Fig. 6(a), are examples of the occurrence of such twists. On the other hand, the also very low frequencies of the longitudinal acoustic modes indicate that compressing or stretching the 3H chains also costs lit-tle energy. These characteristics are compatible with the non-monotonic dependence reported here between encapsulating UNSWCNT diameter and encapsulated coil pitch: Since only small amounts of energy are needed for the Te 3H chains to be twisted, compressed and stretched, the chains can easily adapt to the periodicity and chirality of the encapsulating UNSWCNTs.

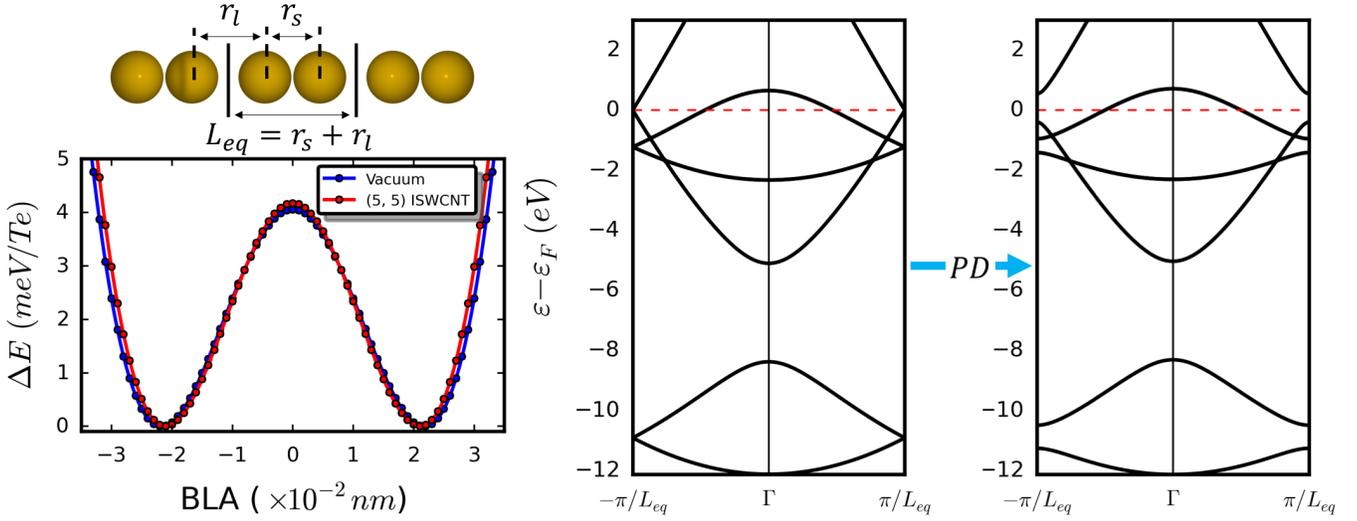

Figure 4: **Total energy shifts** ($\Delta E$) **vs BLA for the Te LC:** Chain in vacuum and encapsulated inside a (5, 5) ICNT, with $BLA \equiv r_l - r_s$. In vacuum, the optimal BLA is 0.022 nm, and the excess energy for the BLA = 0 case is 4.06 meV/Te. For the encapsulated LC, the corresponding values are 0.021 nm and 4.17 meV/Te. **Electronic band structures:** Left-hand side: Chain without PD (*i.e.*, BLA = 0). Right-hand side: PD chain with BLA = 0.022 nm. Mind that $L_{eq}$ refers here to the equilibrium lattice constant of the 2-atom PC of the chain featuring a PD.

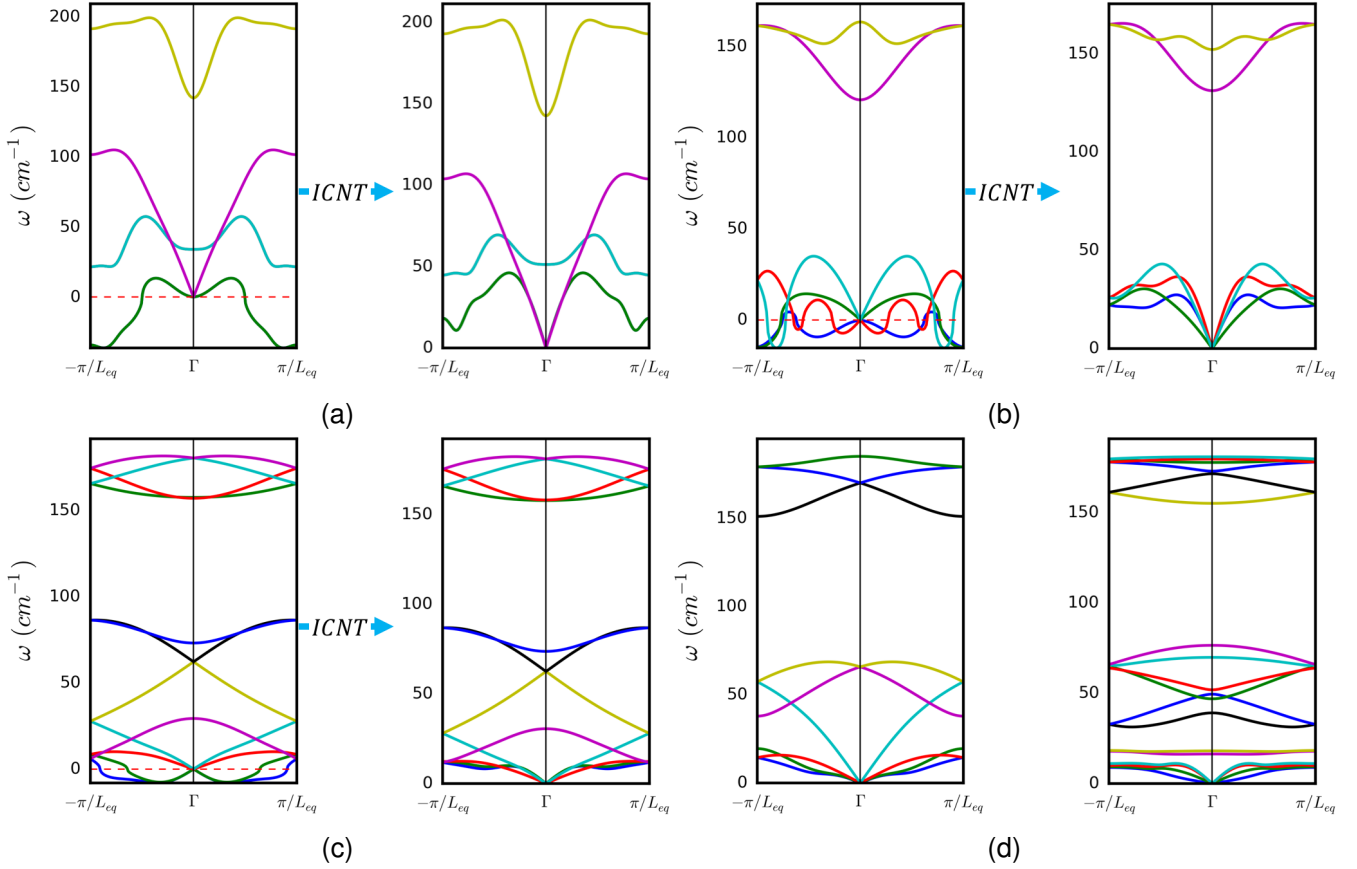

Figure 5: Phonon spectra for the (a) LC, (b) ZZ, (c) 4H, and (d) 3H chains. The figures in (a)-(c) show the spectra calculated for the bare chains (left) and for the chains encapsulated inside (5, 5), (8, 3) and (11, 5) ICNTs, respectively (right). The unstable acoustic modes in vacuum are stabilised upon encapsulation. In (d), the vacuum spectra calculated for the ideal (left) and defective (right) 3H chains are presented. Both 3H chains are mechanically stable.

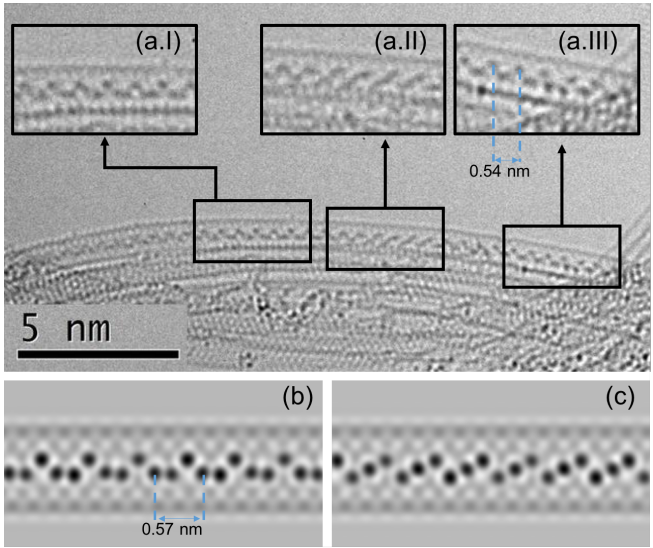

Figure 6: (a) Experimental and (b, c) simulated TEM images for encapsulated Te 3H. The SWCNT diameter in (a) is $0.945 \pm 0.015\,nm$. The simulated images correspond to the most energetically favourable structure found here using AIRSS for Te encapsulated inside an explicit (7, 7) SWCNT ($d = 0.949\,nm$). They differ only by a rotation around the common translation axis. We found that the energy required to compress the AIRSS structure so that theoretical and experimental coil pitches coincide is about $20\,meV$/Te. This is less than $1\,k_BT$/Te at the adopted experimental conditions. The agreement between the geometries shown in (a.I), (a.III) and (b), as well as between (a.II) and (c), is excellent. Simulated images produced using the SimulaTEM code.[40]

## Conclusions

We have reported the observation and modelling of extreme Te nanowires encapsulated inside ultranarrow single-walled carbon nanotubes with diameters between 0.7 and 1.1 $nm$. Using state-of-the-art Aberration Corrected High-Resolution Transmission Electron Microscopy (AC-TEM) and Aberration Corrected Scanning Transmission Electron Microscopy (AC-STEM) methods, we have produced atomic-resolution images of such encapsulated extreme nanowires. Using the AIRSS method, we have successfully predicted the most energetically favourable structures formed by encapsulated Te. We have then produced a diagram of formation energy as function of encapsulating diameters for such structures, providing thus a precise map of the diameter-dependent transitions undergone by the encapsulated extreme nanowires. Such transitions were shown to be electronic (metal-semiconductor) as well as structural. We found a remarkable agreement between our experimental observations and our theoretical modelling, especially with regards to the determination of the encapsulating nanotube diameters at which the structural transitions have been predicted to occur.

We analysed the mechanical stability of the encapsulated Te extreme nanowires and determined whether or not they are expected to form. By doing so, we also demonstrated that nanoconfinement is not just a viable way to synthesise extreme nanowires: In many cases, it might actually be a necessary condition to prevent the nanowires from disintegrating spontaneously. In particular, we have shown that truly 1D, single-atom wide Te linear chains can be synthesised provided that the confining diameters are small enough. Moreover, we established that such chains are longitudinally stabilised by the presence of Peierls structural distortions. We point out that the emergence of linear chains is not an immediate consequence of extreme confinement. In fact, structures such as isolated atoms and dimers, for instance, are also geometrically compatible with the nanoconfinement conditions to which our Te linear chains are submitted, and could thus have been detected/predicted if they were energetically favourable. Finally, for the Te 3-fold-symmetry helical coil, we found that the pitch of the coil varies in a non-monotonic fashion with the encapsulating diameters. Such an effect is related to the relatively low energy cost involved in the processes of compressing, stretching and twisting such coils, which allows them to adapt to the periodicity and chirality of the encapsulating nanotubes.

The excellent agreement between our theoretical modelling and the outcomes of our experiments shows that our experimental-theoretical approach provides a high-quality quantitative characterisation of encapsulated extreme 1D nanowires – in fact, the most precise characterisation of such systems presented to date. We thus anticipate that the present study will set the basic paradigm to be adopted in the characterisation and understanding of nanoconfined 1D materials.

# Experimental Methods

## Sample Preparation

The ENs are prepared by sublimation into pre-treated SWeNT® UNSWCNTs produced by the CoMoCAT process,[41] supplied by Sigma-Aldrich with a stated median diameter range of 0.7-1.1 nm. 100 mg of UNSWCNTs were pre-treated by heating in open air in an alumina boat placed in a Carbolite tube furnace (MTF 12/28/250) to 750 K for 50 minutes, reducing in mass to 60 mg. 100 mg of Te (Sigma-Aldrich, 99.8% trace metals) was then loaded into one end of a sublimation ampule with the 60 mg of UNSWCNT in the other centrally separated by an indentation in the centre of the ampule. The ampule was sealed under vacuum and baked, with the Te section being placed in the centre of the hot zone of the Carbolite tube furnace at 700 K for 6 days. Confirmation of sublimation arose from observations of Te crystallite deposition in the UNSWCNT region of the ampule by TEM. For TEM examination, 3 drops of dispersed CNTs were drop cast onto a 3.05 mm copper grid with lacey carbon support film (Agar Scientific).

## HR-TEM and STEM-EELS

A JEOL ARM 200F microscope operating at 80 kV and equipped with a CEOS aberration corrector and a Gatan SC1000 ORIUS camera with a 4008 2672 pixel CCD was been used for TEM investigations. Wien filtering was performed of selected TEM Images using the script HRTEM FIlter authored by D. R. G. Mitchell and based on the work of R. Kilaas.[42] A Gatan fiber-optical coupled SC1000 ORIUS camera with CCD size of 4008 by 2672 pixel was used for image acquisition. The STEM-EELS data was acquired on a Nion UltraSTEM100 instrument, operated at 60 kV acceleration voltage and equipped with a cold field emission gun providing a native beam energy width of 0.30 eV. The combination of low acceleration voltage and ultra-high vacuum (UHV) conditions at the sample (below 1x10-9 Torr) ensured that damage to the samples, in particular through knock-on or etching, was minimal, enabling long beam dwell times for spectrum imaging. The optics were configured to form a 1.1 Å probe (full-width at half-maximum) of 32 mrad convergence semi-angle, with 40 pA beam current. High and medium angle annular dark field images were recorded with semi-angular ranges of 85-190 and 55-80 mrad, respectively.

EELS data were collected with a Gatan Enfina spectrometer with a collection semi-angle of 36 mrad. A dispersion of 0.5 eV/channel was chosen in order to record simultaneously the C K and Te M4,5 edges (resulting in an effective energy resolution of 1.5 eV, limited by the detector point spread function). Spectrum images were acquired with a dwell time of 0.05s per pixel, providing a good compromise between signal-to-noise and possible damage to the Te wires within the nanotubes. The Te and C maps were generated by integrating the signal above the edge onsets over 75 eV windows, after subtraction of the continuous background using a power law. Principal Component Analysis was carefully applied to the raw datasets to remove detector Poisson noise.[43]

**Acknowledgement** PVCM, AJM, AV and DQ thank the UK Engineering and Physical Sciences Research Council (EPSRC) for funding through grant numbers EP/M011925/1 and EP/M010643/1. JMW acknowledges financial support from the EPSRC Centre for Doctoral Training in Computational Methods for Materials Science under grant EP/L015552/1. JS and SM are further indebted to support from EP/I033394/1. This work used the ARCHER UK National Supercomputing Service (www.archer.ac.uk). Data used in this work are available *via* the Cambridge data repository at doi.org/10.17863/CAM.7096.

**Supporting Information Available:** More information about the adopted conventions,[44,45] terminology and calculation[46–53] parameters are available in the enclosed SI. A detailed discussion about the ICNT model introduced here, as well as about our adaptation of AIRSS for 1D systems (including corrections of mismatch-induced unphysical strains), can also be found there. The SI also presents a general method to calculate the unstrained periodicity of encapsulated nanowires while fully accounting for the presence of the encapsulating (explicit) SWCNT.

**NB:** This manuscript is a preprint of the paper *ACS Nano* **2017**, **DOI:** 10.1021/acsnano.7b02225. When referring to the work presented here, please cite the published *ACS Nano* paper instead.

# References

1. Spencer, J. H.; Nesbitt, J. M.; Trewhitt, H.; Kashtiban, R. J.; Bell, G.; Ivanov, V. G.; Faulques, E.; Sloan, J.; Smith, D. C. Raman Spectroscopy of Optical Transitions and Vibrational Energies of ~1 nm HgTe Extreme Nanowires within Single Walled Carbon Nanotubes. *ACS Nano* **2014**, *8*, 9044–9052.

2. *Carbon Nanotubes: Quantum Cylinders of Graphene*; Elsevier B.V, 2008.

3. Massa, M. V.; Carvalho, J. L.; Dalnoki-Veress, K. Confinement Effects in Polymer Crystal Nucleation from the Bulk to Few-Chain Systems. *Phys. Rev. Lett.* **2006**, *97*, 247802.

4. Carter, R.; Sloan, J.; Kirkland, A. I.; Meyer, R. R.; Lindan, P. J. D.; Lin, G.; Green, M. L. H.; Vlandas, A.; Hutchison, J. L.; Harding, J. Correlation of Structural and Electronic Properties in a New Low-Dimensional Form of Mercury Telluride. *Phys. Rev. Lett.* **2006**, *96*, 215501.

5. Uemura, T.; Horike, S.; Kitagawa, K.; Mizuno, M.; Endo, K.; Bracco, S.; Comotti, A.; Sozzani, P.; Nagaoka, M.; Kitagawa, S. Conformation and Molecular Dynamics of Single Polystyrene Chain Confined in Coordination Nanospace. *J. Am. Chem. Soc.* **2008**, *130*, 6781–6788.

6. Eliseev, A.; Yashina, L.; Kharlamova, M.; Kiselev, N. *Electronic Properties of Carbon Nanotubes*; InTech, 2011; Chapter 8, pp 127–156.

7. Ivanov, A. S.; Kar, T.; Boldyrev, A. I. Nanoscale Stabilization of Zintl Compounds: 1D Ionic Li-P Double Helix Confined inside a Carbon Nanotube. *Nanoscale* **2016**, *8*, 3454–3460.

8. Vasylenko, A.; Wynn, J.; Medeiros, P. V. C.; Morris, A. J.; Sloan, J.; Quigley, D. Encapsulated Nanowires: Boosting Electronic Transport in Carbon Nanotubes. *Phys. Rev. B: Condens. Matter Mater. Phys.* **2017**, *95*, 121408(R).

9. Desai, S. B.; Madhvapathy, S. R.; Sachid, A. B.; Llinas, J. P.; Wang, Q.; Ahn, G. H.; Pitner, G.; Kim, M. J.; Bokor, J.; Hu, C.; Wong, H.-S. P.; Javey, A. MoS2 Transistors with 1-Nanometer Gate Lengths. *Science* **2016**, *354*, 99–102.

10. Gao, Y.; Bando, Y. Carbon Nanothermometer Containing Gallium. *Nature* **2002**, *415*, 599–599.

11. Winkler, A.; Mühl, T.; Menzel, S.; Kozhuharova-Koseva, R.; Hampel, S.; Leonhardt, A.; Büchner, B. Magnetic Force Microscopy Sensors Using Iron-Filled Carbon Nanotubes. *J. Appl. Phys. (Melville, NY, U. S.)* **2006**, *99*, 104905.

12. Gautam, U. K.; Costa, P. M. F. J.; Bando, Y.; Fang, X.; Li, L.; Imura, M.; Golberg, D. Recent Developments in Inorganically Filled Carbon Nanotubes: Successes and Challenges. *Sci. Technol. Adv. Mater.* **2010**, *11*, 054501.

13. Hang, B. T.; Hayashi, H.; Yoon, S.-H.; Okada, S.; Yamaki, J.-i. Fe2O3-Filled Carbon Nanotubes as a Negative Electrode for an Fe-Air Battery. *J. Power Sources* **2008**, *178*, 393–401.

14. Rossella, F.; Soldano, C.; Bellani, V.; Tommasini, M. Metal-Filled Carbon Nanotubes as a Novel Class of Photothermal Nanomaterials.

*Adv. Mater. (Weinheim, Ger.)* **2012**, *24*, 2453–2458.

15. Wuttig, M.; Yamada, N. Phase-Change Materials for Rewriteable Data Storage. *Nat. Mater.* **2007**, *6*, 824–832.

16. Giusca, C. E.; Stolojan, V.; Sloan, J.; Börrnert, F.; Shiozawa, H.; Sader, K.; Rümmeli, M. H.; Büchner, B.; Silva, S. R. P. Confined Crystals of the Smallest Phase-Change Material. *Nano Lett.* **2013**, *13*, 4020–4027.

17. Liu, Z.; Li, S.; Yang, Y.; Hu, Z.; Peng, S.; Liang, J.; Qian, Y. Shape-Controlled Synthesis and Growth Mechanism of One-Dimensional Nanostructures of Trigonal Tellurium. *New J. Chem.* **2003**, *27*, 1748–1752.

18. Gautam, U. K.; Rao, C. N. R. Controlled Synthesis of Crystalline Tellurium Nanorods, Nanowires, Nanobelts and Related Structures by a Self-Seeding Solution Process. *J. Mater. Chem.* **2004**, *14*, 2530–2535.

19. Liang, F.; Qian, H. Synthesis of Tellurium Nanowires and Their Transport Property. *Mater. Chem. Phys.* **2009**, *113*, 523–526.

20. Kudryavtsev, A. A. *The Chemistry and Technology of Selenium and Tellurium*; Collet's Ltd.: London, 1974.

21. Tsiulyanu, D.; Tsiulyanu, A.; Liess, H.-D.; Eisele, I. Characterization of Tellurium-Based Films for $NO_2$ Detection. *Thin Solid Films* **2005**, *485*, 252–256.

22. Miura, N. Fast Photoconductivity of Tellurium by $CO_2$ Laser Radiation. *Appl. Phys. Lett.* **1968**, *12*, 374–375.

23. Liang, W.; Sai-Ling, H.; Fei, Z. Band Structures of Two-Dimensional Photonic Crystals with Regular Polygon Cylinders Calculated by Linear Operations. *Acta Phys. Sin.* **2002**, *51*, 2865–2870.

24. Kobayashi, K.; Yasuda, H. Structural Transition of Tellurium Encapsulated in Confined One-Dimensional Nanospaces Depending on the Diameter. *Chem. Phys. Lett.* **2015**, *634*, 60–65.

25. Suenaga, K.; Sato, Y.; Liu, Z.; Kataura, H.; Okazaki, T.; Kimoto, K.; Sawada, H.; Sasaki, T.; Omoto, K.; Tomita, T.; Kaneyama, T.; Kondo, Y. Visualizing and Identifying Single Atoms Using Electron Energy-Loss Spectroscopy with Low Accelerating Voltage. *Nat. Chem.* **2009**, *1*, 415–418.

26. Krivanek, O. L.; Chisholm, M. F.; Nicolosi, V.; Pennycook, T. J.; Corbin, G. J.; Dellby, N.; Murfitt, M. F.; Own, C. S.; Szilagyi, Z. S.; Oxley, M. P.; Pantelides, S. T.; Pennycook, S. J. Atom-by-Atom Structural and Chemical Analysis by Annular Dark-Field Electron Microscopy. *Nature* **2010**, *464*, 571–574.

27. Pickard, C. J.; Needs, R. J. *Ab Initio* Random Structure Searching. *J. Phys.: Condens. Matter* **2011**, *23*, 053201.

28. Morris, A. J.; Pickard, C. J.; Needs, R. J. Hydrogen/silicon Complexes in Silicon from Computational Searches. *Phys. Rev. B: Condens. Matter Mater. Phys.* **2008**, *78*, 184102.

29. Morris, A. J.; Grey, C. P.; Needs, R. J.; Pickard, C. J. Energetics of Hydrogen/lithium Complexes in Silicon Analyzed Using the Maxwell Construction. *Phys. Rev. B: Condens. Matter Mater. Phys.* **2011**, *84*, 224106.

30. Morris, A. J.; Needs, R. J.; Salager, E.; Grey, C. P.; Pickard, C. J. Lithiation of Silicon *via* Lithium Zintl-Defect Complexes from First Principles. *Phys. Rev. B: Condens. Matter Mater. Phys.* **2013**, *87*, 174108.

31. Pickard, C. J.; Needs, R. J. Highly Compressed Ammonia Forms an Ionic Crystal. *Nat. Mater.* **2008**, *7*, 775–779.

32. Ogata, K.; Salager, E.; Kerr, C.; Fraser, A.; Ducati, C.; Morris, A.; Hofmann, S.; Grey, C. Revealing Lithium-Silicide Phase Transformations in Nano-Structured Silicon-Based Lithium Ion Batteries *via in Situ* NMR Spectroscopy. *Nat. Commun.* **2014**, *5*, 3217.

33. Senga, R.; Komsa, H.-P.; Liu, Z.; Hirose-Takai, K.; Krasheninnikov, A. V.; Suenaga, K.

Atomic Structure and Dynamic Behaviour of Truly One-Dimensional Ionic Chains inside Carbon Nanotubes. *Nat. Mater.* **2014**, *13*, 1050–1054.

34. Nicholls, R. J.; Morris, A. J.; Pickard, C. J.; Yates, J. R. OptaDOS - a New Tool for EELS Calculations. *J. Phys.: Conf. Ser.* **2012**, *371*, 012062.

35. Pouget, J. P. The Peierls Instability and Charge Density Wave in One-Dimensional Electronic Conductors. *C. R. Phys.* **2016**, *17*, 332–356.

36. Batra, I. P. Gapless Peierls Transition. *Phys. Rev. B: Condens. Matter Mater. Phys.* **1990**, *42*, 9162–9165.

37. Voit, J. Dynamical Correlation Functions of One-Dimensional Superconductors and Peierls and Mott Insulators. *Eur. Phys. J.* **1998**, *5*, 505–519.

38. Ghosh, P.; Kahaly, M.; Waghmare, U. Atomic and Electronic Structures, Elastic Properties, and Optical Conductivity of Bulk Te and Te Nanowires: A First-Principles Study. *Phys. Rev. B: Condens. Matter Mater. Phys.* **2007**, *75*, 245437.

39. Li, G.; Fu, C.; Oviedo, M. B.; Chen, M.; Tian, X.; Bekyarova, E.; Itkis, M. E.; Wong, B. M.; Guo, J.; Haddon, R. C. Giant Raman Response to the Encapsulation of Sulfur in Narrow Diameter Single-Walled Carbon Nanotubes. *J. Am. Chem. Soc.* **2016**, *138*, 40–43.

40. Gómez-Rodríguez, A.; Beltrán-del Río, L.; Herrera-Becerra, R. SimulaTEM: Multislice Simulations for General Objects. *Ultramicroscopy* **2010**, *110*, 95–104.

41. Jorio, A.; Santos, A. P.; Ribeiro, H. B.; Fantini, C.; Souza, M.; Vieira, J. P. M.; Furtado, C. A.; Jiang, J.; Saito, R.; Balzano, L.; Resasco, D. E.; Pimenta, M. A. Quantifying Carbon-Nanotube Species with Resonance Raman Scattering. *Phys. Rev. B: Condens. Matter Mater. Phys.* **2005**, *72*, 075207.

42. Kilaas, R. Optimal and Near-Optimal Filters in High-Resolution Electron Microscopy. *J. Microsc. (Oxford, U. K.)* **1998**, *190*, 45–51.

43. Watanabe, M.; Okunishi, E.; Ishizuka, K. Analysis of Spectrum-Imaging Datasets in Atomic-Resolution Electron Microscopy. *Microsc. Anal. (U.K. Ed.)* **2009**, *23*, 5–7.

44. Saito, R.; Dresselhaus, G.; Dresselhaus, M. S. *Physical Properties of Carbon Nanotubes*; Imperial College Press: London, 2001.

45. Qin, L.-C. Determination of the Chiral Indices (n,m) of Carbon Nanotubes by Electron Diffraction. *Phys. Chem. Chem. Phys.* **2007**, *9*, 31–48.

46. Hohenberg, P.; Kohn, W. Inhomogeneous Electron Gas. *Phys. Rev.* **1964**, *136*, B864.

47. Kohn, W.; Sham, L. J. Self-Consistent Equations Including Exchange and Correlation Effects. *Phys. Rev.* **1965**, *140*, A1133.

48. Clark, S. J.; Segall, M. D.; Pickard, C. J.; Hasnip, P. J.; Probert, M. I. J.; Refson, K.; Payne, M. C. First Principles Methods Using CASTEP. *Z. Kristallogr. Cryst. Mater.* **2005**, *220*, 567–570.

49. Vanderbilt, D. Soft Self-Consistent Pseudopotentials in a Generalized Eigenvalue Formalism. *Phys. Rev. B: Condens. Matter Mater. Phys.* **1990**, *41*, 7892–7895.

50. Kozinsky, B.; Marzari, N. Static Dielectric Properties of Carbon Nanotubes from First Principles. *Phys. Rev. Lett.* **2006**, *96*, 166801.

51. Arfken, G. B.; Weber, H. J. *Mathematical Methods for Physicists*, 6th ed.; Academic Press: San Diego, 2005.

52. Van Siclen, C. Estimating Atomic 1s Orbital Radii. *J. Comput. Appl. Math.* **1987**, *19*, 283–286.

53. Vinet, P.; Smith, J. R.; Ferrante, J.; Rose, J. H. Temperature Effects on the Universal Equation of State of Solids. *Phys. Rev. B: Condens. Matter Mater. Phys.* **1987**, *35*, 1945–1953.

# Supporting Information for

# "Single-Atom Scale Structural Selectivity in Te Nanowires Encapsulated inside Ultra-Narrow, Single-Walled Carbon Nanotubes"

Paulo V. C. Medeiros,*,† Samuel Marks,‡ Jamie M. Wynn,†
Andrij Vasylenko,‡,¶ Quentin M. Ramasse,§ David Quigley,‡,∥ Jeremy Sloan,*,‡
and Andrew J. Morris*,†,‡

†*Theory of Condensed Matter Group, Cavendish Laboratory, University of Cambridge, J. J. Thomson Avenue, Cambridge CB3 0HE, U.K.*
‡*Department of Physics, University of Warwick, Coventry CV4 7AL, U.K.*
¶*Institute for Condensed Matter Physics, National Academy of Science of Ukraine (NAS Ukraine), 1 Sventsitskii street, 79011 Lviv, Ukraine*
§*SuperSTEM Laboratory, STFC Daresbury, Keckwick Lane, Daresbury WA4 4AD, U.K.*
∥*Centre for Scientific Computing, University of Warwick, Coventry CV4 7AL, U.K.*

E-mail: pvm20@cam.ac.uk; j.sloan@warwick.ac.uk; ajm255@cam.ac.uk

# Contents

# 1 Carbon Nanotubes: Terminology and Conventions

Adopting the convention widely used in the literature, we use a tuple $(n, m)$ of integers to refer to a single-walled carbon nanotube (CNT) (SWCNT) with *chiral vector* $\mathbf{C}_{nm} \equiv n \cdot \mathbf{a}_1 + m \cdot \mathbf{a}_2$, where $\mathbf{a}_1$ and $\mathbf{a}_2$ represent the vectors of the hexagonal primitive unit cell of graphene. An $(n, m)$ SWCNT is generated by rolling up a sheet of graphene so that the points originally located at $\mathbf{r}$ and $\mathbf{r} + \mathbf{C}_{nm}$ coincide after the rolling up process has taken place. The periodicity of a SWCNT is determined by its *translation vector*, $\mathbf{T}_{nm}$, which is parallel to the SWCNT axis and is defined as the smallest vector perpendicular to $\mathbf{C}_{nm}$ in the graphene's lattice. A lattice translation in the originating graphene sheet can always be mapped, in a SWCNT, to a translation along $\mathbf{T}_{nm}$ followed by a rotation around it. Due to the hexagonal symmetry of graphene, all symmetry-unrelated SWCNTs can be generated by choosing $0 \leq |m| \leq n$,[1,2] where the choices of $m > 0$ and $m < 0$ give rise to right- and left-handed SWCNTs, respectively.[2] The tuples $(n > 0, |m| \leq n)$ are therefore unique identifiers for all independent SWCNTs. In this work, $n$ and $m$ lie within the ranges thus specified.

# 2 Theoretical Calculations

## 2.1 Structural Searches – AIRSS, 1D Structures, Mismatch and Strain

In the *ab initio* random structure searching (AIRSS) method,[3] target chemical compositions are taken as the only input information to randomly generate a large number of initial candidate geometries for the studied material. The survey of the search space can be aided by considering physically sensible constraints such as symmetry group considerations, for instance. The geometries thus generated are then fully relaxed into the nearest local minimum in their potential energy surfaces using an *ab initio* method of choice, such as density-functional theory (DFT), as used here. The resulting structures are finally ranked according to some well defined metrics, e.g. formation energy per filling atom. Provided that a large enough number of distinct sensible structures is generated, AIRSS can often unravel the ground state geometries of the surveyed materials, as well many of the most relevant metastable configurations.[3]

A difficulty that arises when doing structural searches on SWCNT-encapsulated nanowires is the fact that the natural periodicity of the encapsulated nanowire (ENW), i.e., the equilibrium length of one repeat unit of the ENW when not constrained to match the encapsulating SWCNT, is generally not commensurate with allowed translations of the encapsulating SWCNT. This means that the ENWs are often subjected to strain in order for matching to be attained. In the best case scenario, such strains affect the energetics of the simulated systems only by the addition an excess strain energy. In more critical situations, however, they can induce spurious local minima in the potential energy landscape into which the structures may be relaxed, leading thus to erroneous structural predictions. Another important observation is that the natural periodicity of an ENW, as well as the dimensions of its lateral cross-section, should depend on the diameters of the confining SWCNTs. For instance, a zigzag chain should eventually turn into a linear structure as it gets squashed by encapsulation inside narrower and narrower SWCNTs. Dealing with such dependences, however, is generally not a straightforward task, specially in high-throughput methods such as AIRSS.

To work around these issues to a first approximation, we have introduced here an electrostatic, smoothed out model for the encapsulating SWCNTs: The *implicit SWCNT (ICNT)* model, discussed in Section 3. We have employed ICNTs to replace the actual SWCNTs during the preliminary stages of our AIRSS searches. After such preliminary AIRSS searches, the most reasonable structures found were further refined and analysed using actual, explicit SWCNTs. Analogously, in the construction of the phase diagram shown in Fig. 1 (main text), we first determined the diameter-dependent lengths and cross-sections of the selected ENWs by performing geometry optimisations inside ICNTs. The resulting structures were then encapsulated inside explicit SWCNTs, subjected to a maximum strain within which the deformations were verified to still be in the elastic regime (typically < 4%). The geometries of ENW+SWCNT systems were then fully optimised before obtaining the formation energies. The use of ICNTs allows a much better control over the effects of strain on the ENWs, and leads to significant speed ups in the searches.

## 2.2 Computational Details

Our *ab initio* DFT[4] calculations were performed within the Kohn–Sham (KS) ensemble[5] and using a plane-wave basis-set for the expansion of the KS states, as implemented in the CASTEP code.[6] We used a variant of version 17 of CASTEP modified by us to allow the use of the ICNTs. Ultrasoft pseudopotentials[7] were employed, and the plane-wave basis set was truncated using a cutoff kinetic energy of 600 eV. The adopted [core].valence electronic configurations for the C and Te atoms were, respectively, $[He].2s^22p^2$ and $[Kr+4d^{10}].5s^25p^4$. In order to estimate the errors introduced by pseudisation of the core, we compared the all-electron total energies of isolated C and Te with the ones calculated using the pseudopotentials as described. We found that the maximum errors in the total energies of C and Te are smaller than 1 meV for the chosen cutoff energy.

The SWCNTs used in our searches were surrounded by a vacuum of 5 Å, which we found to be enough to prevent spurious interactions between neighbouring images of the simulated systems, while allowing the searches to be performed efficiently. The Brillouin zones were sampled using $1 \times 1 \times n$ Monkhorst–Pack grids, where $n$ was chosen, for each $N_{\mathbf{T}_{nm}} \times (n,m)$ SWCNT, so that the distance between neighbouring $k$-points along the reciprocal lattice direction parallel to $\mathbf{T}_{nm}$ was of about $0.036 \times 2\pi$ Å$^{-1}$. We found this to be enough to converge the calculated $E_{Form}^{Te}$ to approximately 10 meV. Phonon calculations were performed using the finite displacement method with $1 \times 1 \times 8$ supercells and a reduced $k$-point spacing of $0.02 \times 2\pi$ Å$^{-1}$ – except for the linear chain (LC) and zigzag configurations, for which the $k$-point spacing was further reduced to $0.0125 \times 2\pi$ Å$^{-1}$. The same $0.0125 \times 2\pi$ Å$^{-1}$ $k$-point spacing was used for the LC Peierls distortion calculations in order to resolve energy differences of the order of 1 meV per atom. We found such a fine spacing to be very important in the detection of Peierls instabilities.

The SWCNTs in our structural searches were randomly selected from a list of all possible symmetry-independent $N_{\mathbf{T}_{nm}} \times (n,m)$ SWCNTs with lengths between 0.5 and 2.2 *nm* and diameters in the range between 0.5 and 1.2 *nm*, where $N_{\mathbf{T}_{nm}}$ represents the number of occurrences of the primitive unit cell of the $(n,m)$ SWCNT along the translation vector $\mathbf{T}_{nm}$. The selected SWCNTs were then filled with Te atoms, as described in Section 2.1, and the initial structures were allowed to relax fully until the maximum norm of the residual Hellmann-Feynman forces acting upon the atoms were not larger than $5 \times 10^{-2} eV/\text{Å}$ for the first screening, and $1 \times 10^{-2} eV/\text{Å}$ during the refinement stage and for the construction of the phase diagram. The maximum allowed residual force norm for the phonon and calculations Peierls distortion calculations was $1 \times 10^{-3} eV/\text{Å}$.

The ranking of the geometry-optimised structures is based on the values of the formation energies per Te atom, $E_{Form}^{Te}$, calculated for each final structure as:

$$E_{Form}^{Te} \equiv \frac{1}{N_{Te}} E_{Form} = \frac{1}{N_{Te}} \left[ E_T - N_C \mu_C - N_{Te} \mu_{Te} \right]. \quad \text{(S1)}$$

In Eq. (S1), $E_T$ is the total energy of the full Te+SWCNT system, $N_C$ and $N_{Te}$ represent the total number of C atoms in the SWCNTs and Te atoms encapsulated inside them, and $\mu_C$ and $\mu_{Te}$ are the chemical potentials associated with a carbon atom in the selected SWCNT and an atom in the bulk phase of Te. To correct for the excess energy present in the encapsulated systems due to the axial strain imposed onto the ENW geometries, a slight modification was made to Eq. (S1) when constructing the phase diagram shown in Fig. 1 (main text). The strain-corrected formation energies were calculated as:

$$E_{Form,PhaseDiag}^{Te} = \frac{1}{N_{Te}} \left[ E_T - N_C \mu_C - N_{Te} \mu_{Te} - E_{Strain} \right], \quad \text{(S2)}$$

where $E_{Strain}$ represents the excess strain energy. We calculated $E_{Strain}$ by first replacing the SWCNTs in the encapsulated systems by the corresponding ICNTs. Then, we allowed the system to relax fully while keeping the unit cell vectors fixed (i.e., without altering the lengths of the ENWs). The resulting systems were then used as starting points for new geometry optimisations inside the same ICNTs, but this time allowing for variations in the length of the ENWs. Finally, the corrections $E_{Strain}$ were obtained by calculating the differences between the energies of the constrained and unconstrained ICNT-encapsulated ENW systems.

# 3  Implicit Carbon Nanotubes

ICNTs are introduced in our calculations by adding external local cylindrically-symmetric confining potentials of the form

$$\phi_{r_0}^{Imp}(r) = \phi_0 \left[ e^{-k_0^2 \cdot (r-r_0)^2} + e^{-k_0^2 \cdot (r+r_0)^2} \right] \tag{S3}$$

to the DFT Hamiltonian of the simulated systems. In Eq. (S3), $r_0$ is the radius of the ICNT – which can, in principle, differ from the radius of the actual SWCNT it represents (see Section 3.1), $r$ is the perpendicular distance between a point in space and the axis of the ICNT, $\phi_0 = \phi_0(r_0) > 0$, and $k_0 = k(r_0) \neq 0$ is a real number. The term $e^{-k_0^2 \cdot (r-r_0)^2}$ in Eq. (S3) reaches its maximum value at the walls of the ICNT. The term $e^{-k_0^2 \cdot (r+r_0)^2}$ is mostly negligible when compared with $e^{-k_0^2 \cdot (r-r_0)^2}$, but its inclusion ensures that the gradient of $\phi_{r_0}^{Imp}(r)$ vanishes identically for points at the symmetry axis of the ICNT ($r = 0$).

## 3.1  Determining the Implicit SWCNT Parameters

In order to determine the values of the parameters $r_0$, $\phi_0$ and $k_0$ in Eq. (S3), we first observe that $\phi_{r_0}^{Imp}(r)$ can be associated to an implicit charge density distribution $\rho_{r_0}^{Imp}(r)$ through the use of Poisson's equation:

$$\nabla^2 \phi_{r_0}^{Imp}(r) = \frac{1}{r}\frac{d}{dr}\left[ r \frac{d}{dr} \phi_{r_0}^{Imp}(r) \right] = -\frac{\rho_{r_0}^{Imp}(r)}{\varepsilon_{r_0}^\perp}, \tag{S4}$$

where $\varepsilon_{r_0}^\perp$ is the transverse static permittivity of the ICNT, and we have used the fact that $\phi_{r_0}^{Imp}(r)$ depends only on the cylindrical radial coordinate $r$ measured from the axis of the ICNT. By combining Eqs. (S3) and (S4) one can easily verify that, for any valid values of $r_0, \phi_0$ and $k_0$,

$$\int_0^\infty \rho_{r_0}^{Imp}(r)\, r\, dr = 0. \tag{S5}$$

Therefore, the total charge associated to $\rho_{r_0}^{Imp}(r)$, and thus to $\phi_{r_0}^{Imp}(r)$, is always null. The explicit form of $\rho_{r_0}^{Imp}(r)$ is also immediately obtained from Eqs. (S3) and (S4):

$$\rho_{r_0}^{Imp}(r) = 2k_0^2 \phi_0 \varepsilon_{r_0}^{\perp} \left\{ \left[ 2 - \frac{r_0}{r} - 2k_0^2(r-r_0)^2 \right] e^{-k_0^2(r-r_0)^2} + \left[ 2 + \frac{r_0}{r} - 2k_0^2(r+r_0)^2 \right] e^{-k_0^2(r+r_0)^2} \right\}. \quad (S6)$$

The general shape of the implicit charge density represented by Eq. (S6) is shown in Fig. S1.

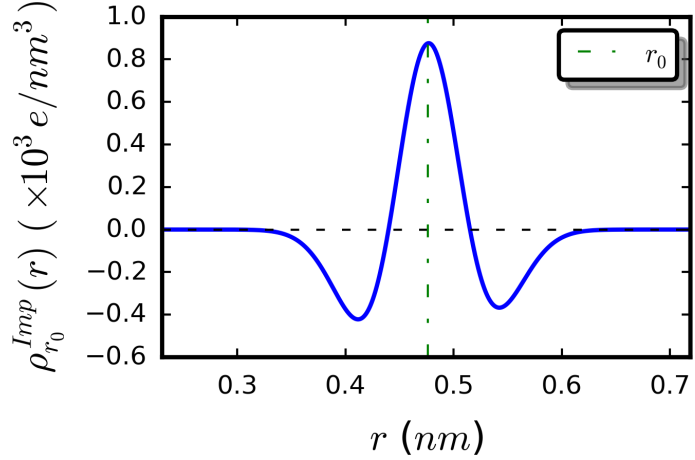

Figure S1: General shape of the total charge density distribution associated to the potential of an ICNT, as defined in Eq. (S3) through the use of Poisson's equation. Positive and negative values of $\rho_{r_0}^{Imp}(r)$ indicate regions of lack and accumulation of electrons, respectively. The total charge associated with $\rho_{r_0}^{Imp}(r)$ is always null, as discussed in the text.

To obtain $\varepsilon_{r_0}^{\perp}$, we note that, as discussed in Ref. [8], the single-tube transverse polarisability per unit length, $\alpha_{r_0}^{\perp}$, relates to $\varepsilon_{r_0}^{\perp}$ as

$$\alpha_{r_0}^{\perp} = \frac{\Omega}{2\pi} \frac{\varepsilon_{r_0}^{\perp} - \varepsilon_0}{\varepsilon_{r_0}^{\perp} + \varepsilon_0} \approx C \cdot (r_0 + \Delta r_0)^2, \quad (S7)$$

from which it follows that

$$\varepsilon_{r_0}^{\perp} = \frac{1 + \frac{2\pi C \cdot (r_0 + \Delta r_0)^2}{\Omega}}{1 - \frac{2\pi C \cdot (r_0 + \Delta r_0)^2}{\Omega}} \varepsilon_0, \quad (S8)$$

where $\varepsilon_0$ is the vacuum permittivity, and $\Omega$ is the area of the section of the plan perpendicular to the axis of the SWCNT contained inside the simulation supercell, while and $C$ and $\Delta r_0$ are positive constants. By explicitly calculating $\alpha_{r_0}^{\perp}$ for a series of chiral and non-chiral SWCNTs, we have found $C \approx 0.039$ and $\Delta r_0 \approx 0.147 nm$, in reasonable agreement with the values reported in Ref. [8]. Eq. (S6) thus becomes:

$$\rho_{r_0}^{Imp}(r) = \frac{1 + \frac{2\pi C \cdot (r_0+\Delta r_0)^2}{\Omega}}{1 - \frac{2\pi C \cdot (r_0+\Delta r_0)^2}{\Omega}} 2k_0^2 \phi_0 \varepsilon_0 \cdot \left\{ \left[ 2 - \frac{r_0}{r} - 2k_0^2(r-r_0)^2 \right] e^{-k_0^2(r-r_0)^2} + \left[ 2 + \frac{r_0}{r} - 2k_0^2(r+r_0)^2 \right] e^{-k_0^2(r+r_0)^2} \right\}. \quad (S9)$$

Next, we calculate the total valence charges $\rho_v(r, \theta, z)$ of the actual SWCNTs. These charges are originally stored in a cartesian $(x_i, y_j, z_k)$ grid, which we then transform into a cylindrical $(r_i, \theta_j, z_k)$ grid in order to evaluate, numerically, the cylindrically-averaged valence charge densities $\tilde{\rho}_v(r)$ defined as:

$$\tilde{\rho}_v(r) \equiv \lim_{\delta r \to 0} \frac{\int_{r'=r-\frac{\delta r}{2}}^{r+\frac{\delta r}{2}} \int_{\theta=-\pi}^{\pi} \int_{z=0}^{L} \rho_v(r', \theta, z) r' dz d\theta dr'}{\int_{r'=r-\frac{\delta r}{2}}^{r+\frac{\delta r}{2}} \int_{\theta=-\pi}^{\pi} \int_{z=0}^{L} r' dz d\theta dr'}, \quad (S10)$$

where $L$ is the length of the smallest translation vector of the SWCNT. Since the integrals of $\tilde{\rho}_v(r)$ and $\rho_v(r, \theta, z)$ over the entire space should be identical, the following normalisation condition holds:

$$\int_0^\infty \tilde{\rho}_v(r) r \, dr = -\frac{Q_v}{2\pi L} = -\frac{4q_v}{a_0^2 \sqrt{3}} r_0, \quad (S11)$$

where $q_v$ and $Q_v$ represent, respectively, the absolute values of the valence charge per carbon atom and of the total valence charge contained in one unit cell of the SWCNT. Interestingly, we have found that, regardless of the SWCNT, $\tilde{\rho}_v(r)$ can be very well approximated by Gaussian distributions centred around the radii of the respective SWCNTs, all having approximately the same standard deviation ($\sim 0.06 nm$). Assuming thus a Gaussian functional form for $\tilde{\rho}_v(r)$ and normalising it according to Eq. (S11), we obtain:

$$\tilde{\rho}_v(r) = -\frac{8q_v k_v}{a_0^2 \sqrt{3\pi} \left[ 2 - \mathbb{Q}(-\frac{1}{2}, k_v^2 r_0^2) \right]} e^{-k_v^2(r-r_0)^2}, \quad (S12)$$

with $\mathbb{Q}(a, z)$ representing the regularised incomplete Gamma function, defined as:[9]

$$\mathbb{Q}(a, z) \equiv \frac{\Gamma(a, z)}{\Gamma(a)}, \tag{S13}$$

where $\Gamma(a, z)$ is the upper incomplete Gamma function,

$$\Gamma(a, z) \equiv \int_z^\infty t^{a-1} e^{-t} dt, \tag{S14}$$

and $\Gamma(a) = \Gamma(a, 0)$ is the ordinary Gamma function. Examples of such $\tilde{\rho}_v(r)$ distributions are shown in Fig. S2.

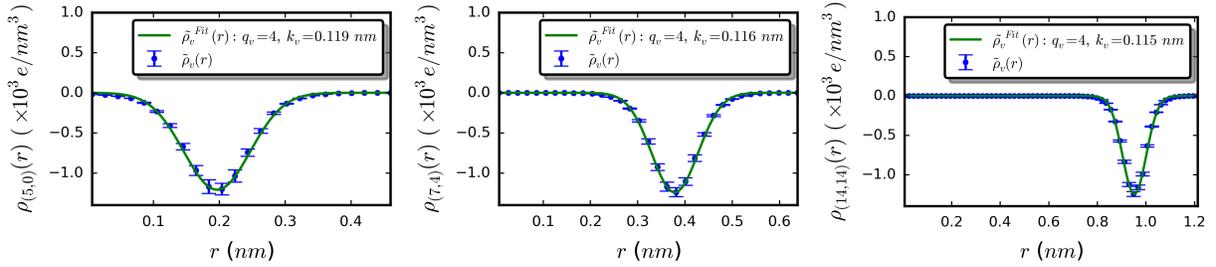

Figure S2: Cylindrically-averaged valence charge density distributions for the (5, 0), (7, 4) and (14, 14) SWCNTs. The variable $r$ denotes the radial distance measured from the centre of the SWCNT, and the vertical error bars indicate the mean square deviations of the average values.

Finally, we observe that, since the core charges of the carbon atoms in the SWCNTs are confined to a very narrow region around the nuclei (the average $1s$ electron radius for carbon is $\sim 0.009 nm$[10]), then, for $|r - r_0| > r_c$, where $r_c$ is the core radius of the carbon pseudopotential used, the total charge density should, for our practical purposes, coincide with the valence one. The final step in our procedure is then to fit the tail of the charge density $\rho_{r_0}^{Imp}(r)$ from Eq. (S9), i.e., the part of $\rho_{r_0}^{Imp}(r)$ located in the region $|r - r_0| > r_c$, to the averaged valence charge $\tilde{\rho}_v(r)$ defined in Eq. (S12), determining thus the values of the parameters $r_0$, $\phi_0$ and $k_0$ in Eq. (S3). An illustration of such a procedure is shown in Fig. S3.

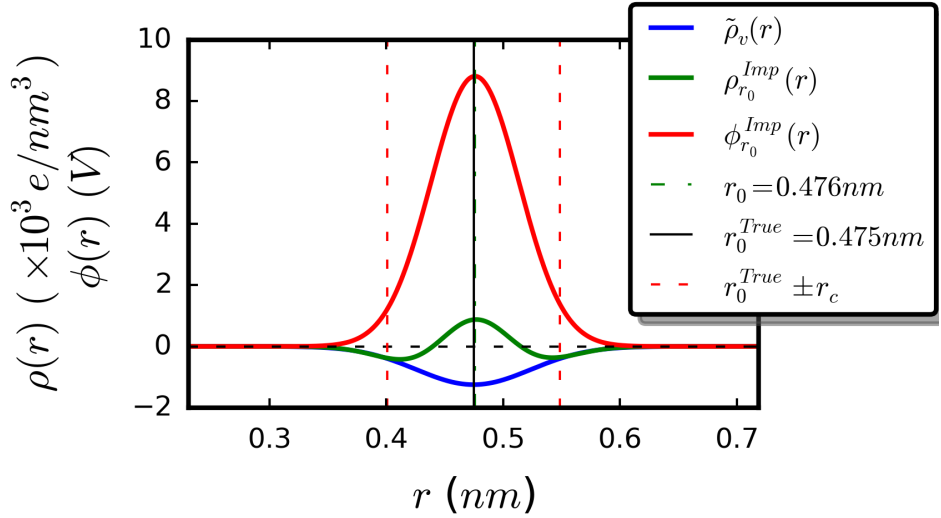

Figure S3: Parameter fitting scheme employed to determine the parameters defining the ICNT confining potentials: An implicit (7, 7) SWCNT.

## 4 Periodicity of Encapsulated Nanowires: Explicit CNT Approach

Although ENW are one-dimensional structures when considered from a physical point of view, one can think of their constituent atoms as occupying a finite, albeit very small, cross-sectional area. It is easy to verify in calculations, for instance, that the atoms in an ENW tend to get closer to the ENW principal axis as the ENW is stretched, while they tend to get farther away from the same axis as the ENW is compressed. Moreover, taking into account the finite dimensions of any given atom, even an atomic linear chain can be thought to possess a finite cross-sectional area determined by the radii of the atoms that form it. Here, we define the volume $V_{ENW}$ of one primitive unit cell (PC) of an ENW (possibly subjected to some strain) as:

$$V_{ENW} \equiv \frac{1}{4}\pi L \cdot D^2, \qquad (S15)$$

where $L$ is the length of the ENW PC and $D$ is the diameter of the smallest cylinder needed to encapsulate the ENW, taking into account the van der Waals radii of its composing atoms. With $V_{ENW}$ defined as in Eq. (S15), we fit a set of calculated $(V_i, E_i)$ points to a theoretical $E(V)$ relation, obtained using an equation of state (EOS), to obtain $V_{eq}$.

As already pointed out, however, $D$ will generally be a function of $L$. Therefore, one needs to know $D(L)$ to work out the ENW equilibrium length $L_{eq}$ from $V_{eq}$. The quantity relating $D$ and $L$ is the Poisson's ratio ($\nu$), defined as:

$$\nu \equiv -\frac{\varepsilon_\perp}{\varepsilon_z}, \tag{S16}$$

where $\varepsilon_\perp$ and $\varepsilon_z$ are, respectively, the strains on the ENW along its transverse and longitudinal directions (not to be confused with the transverse static permittivity of the ICNTs introduced in Eq. (S4)):

$$\varepsilon_z = \frac{\Delta L}{L_{eq}} = \frac{L - L_{eq}}{L_{eq}}, \quad \varepsilon_\perp = \frac{\Delta D}{D_{eq}} = \frac{D - D_{eq}}{D_{eq}}, \tag{S17}$$

which are here assumed to be small. It follows from Eqs. (S16) and (S17) that:

$$\nu = -\frac{L_{eq}}{D_{eq}} \frac{\Delta D}{\Delta L}. \tag{S18}$$

Since $L_{eq}$ and $D_{eq}$ are both unknown quantities, it is useful investigating the limit of infinitesimal deformations of the ENW PC around the equilibrium configuration. This can be done by writing

$$\nu = -\lim_{\delta L \to 0} \frac{L - \delta L}{D - \delta D} \frac{D - (D - \delta D)}{L - (L - \delta L)}, \tag{S19}$$

where $L = L_{eq} + \delta L$ and $D = D_{eq} + \delta D(L)$. Since $\delta D(L) \to 0$ as $\delta L \to 0$, Eq. (S19) thus leads to:

$$\nu = -\frac{L}{D} \frac{dD}{dL}. \tag{S20}$$

Therefore, if we choose an arbitrary reference point $(L_0, D_0)$ located in the vicinity of $(L_{eq}, D_{eq})$, then a good approximation for the diameter $D(L)$, where $L$ is also located close to $L_{eq}$, can be obtained by integrating Eq. (S20):

$$D(L) = D_0 \left(1 + \frac{L - L_0}{L_0}\right)^{-\nu}. \tag{S21}$$

We determine the value of $\nu$ by performing a series of geometry optimisations for the ENW kept at different fixed lengths $L$, calculating $D$ for each relaxed structure as discussed, and then fitting Eq. (S21) through the points thereby obtained. $L_0$ is kept fixed during the fit, while $D_0$ and $\nu$ are allowed to vary. After using an EOS to calculate $V_{eq}$, we substitute Eq. (S21) into Eq. (S15) and invert the resulting equation to determine $L_{eq}$. Now, since, ideally, one would like to have $L_0 = L_{eq}$ and $D_0 = D_{eq}$ in Eq. (S21), we repeat the process just described, self-consistently, making $L_0^{(i)} \to L_{eq}^{(i-1)}$. Convergence is assumed to have been achieved when $\left|L_{eq}^{(i)} - L_0^{(i)}\right| < 10^{-11}$ nm, $\left|D_0^{(i)} - \sqrt{4V_{eq}^{(i)}/(\pi L_{eq}^{(i)})}\right| < 10^{-11}$ nm, and $\left|\nu^i - \nu^{i-1}\right| < 10^{-10}$.

## 4.1 Non-Monotonic $L_{eq}(d_{CNT})$ Relation

We have applied the above described procedure to determine $L_{eq}$ for 3-fold-symmetry helical coil (3H) nanowires encapsulated inside ultra-narrow single-walled carbon nanotubes (UNSWCNTs) with diameters $d_{CNT}$ between 0.84 and 0.96 nm. The values of $L$ picked for the ENW PCs throughout the calculations described in this section were chosen as to lie approximately within ±10% of the equilibrium PC lengths estimated, to a first approximation, using ICNTs (see Section 3). The Rose-Vinet EOS[11] was adopted to obtain $E(V)$.

In order to account for the presence of actual, explicit CNTs in the ENW systems, the calculated formation energies per PC, defined similarly to Eq. (S1), were used instead of the total energies. Fig. S4 illustrates the fitting procedure, and Fig. S5 shows a comparison between experimentally measured and theoretically calculated 3H ENW lengths as a function of the encapsulating SWCNTs. Both the experimental and theoretical curves show a non-monotonic relation between $L_{eq}$ and the diameters of the encapsulating UNSWCNT.

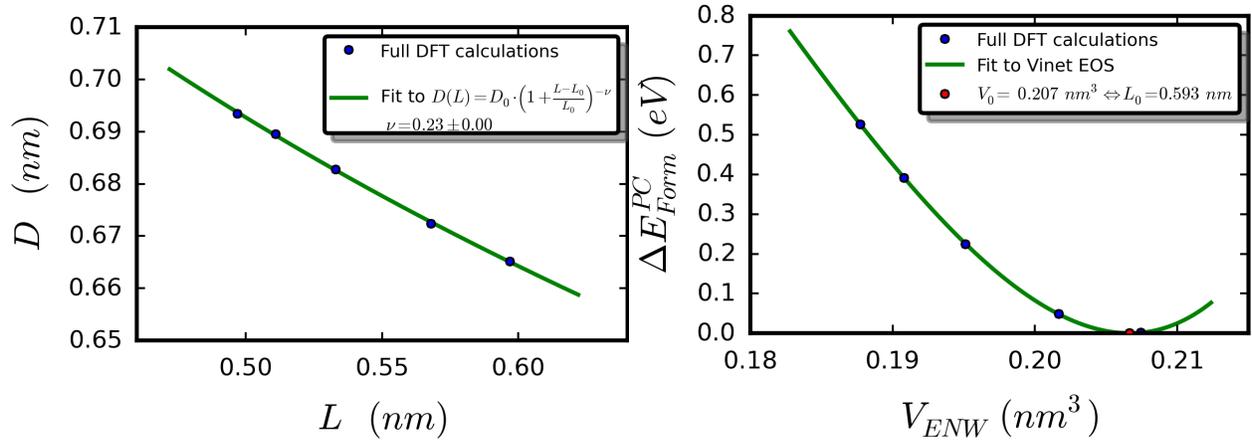

Figure S4: 3H chain encapsulated inside an explicit (11, 0) SWCNT. Left – determining $\nu$: ENW diameter ($D$) as a function of the ENW PC length (i.e., 3H coil pitch, $L$), calculated as described in the text, along with a fit to Eq. (S21). Right – determining $V_{eq}$ and $L_{eq}$: Shifts in the formation energies ($\Delta E_{Form}^{PC}$, with respect to the calculated minimum one) as a function of the ENW volume ($V_{ENW}$), calculated from D and L using Eq. (S15), along with a fit to the Rose-Vinet EOS.

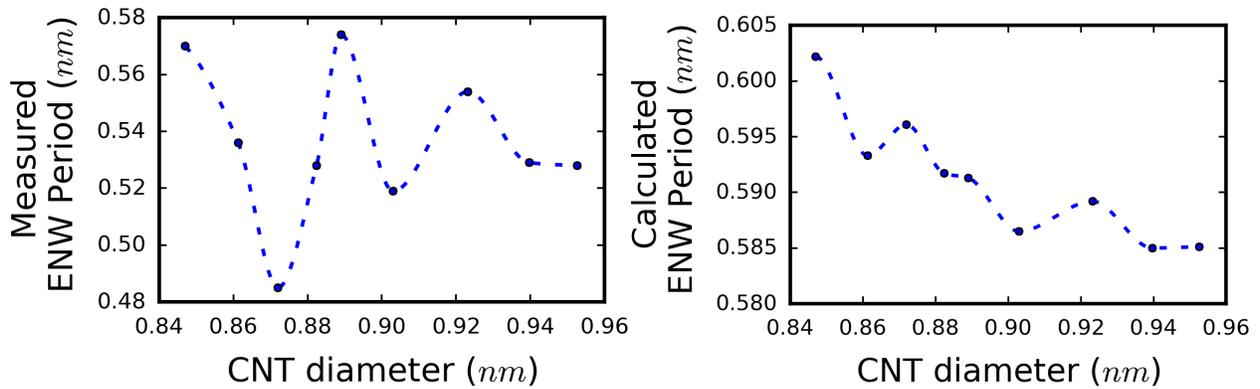

Figure S5: Encapsulated 3H chain: 3H coil pitch as a function of the diameters of the encapsulating UNSWCNTs. Experimental data and theoretical results are shown on the left- and right-hand sides, respectively.

# References

1. Saito, R.; Dresselhaus, G.; Dresselhaus, M. S. *Physical Properties of Carbon Nanotubes*; Imperial College Press: London, 2001.

2. Qin, L.-C. Determination of the Chiral Indices (n,m) of Carbon Nanotubes by Electron Diffraction. *Phys. Chem. Chem. Phys.* **2007**, *9*, 31–48.

3. Pickard, C. J.; Needs, R. J. *Ab Initio* Random Structure Searching. *J. Phys.: Condens. Matter* **2011**, *23*, 053201.

4. Hohenberg, P.; Kohn, W. Inhomogeneous Electron Gas. *Phys. Rev.* **1964**, *136*, B864.

5. Kohn, W.; Sham, L. J. Self-Consistent Equations Including Exchange and Correlation Effects. *Phys. Rev.* **1965**, *140*, A1133.

6. Clark, S. J.; Segall, M. D.; Pickard, C. J.; Hasnip, P. J.; Probert, M. I. J.; Refson, K.; Payne, M. C. First Principles Methods Using CASTEP. *Z. Kristallogr. Cryst. Mater.* **2005**, *220*, 567–570.

7. Vanderbilt, D. Soft Self-Consistent Pseudopotentials in a Generalized Eigenvalue Formalism. *Phys. Rev. B: Condens. Matter Mater. Phys.* **1990**, *41*, 7892–7895.

8. Kozinsky, B.; Marzari, N. Static Dielectric Properties of Carbon Nanotubes from First Principles. *Phys. Rev. Lett.* **2006**, *96*, 166801.

9. Arfken, G. B.; Weber, H. J. *Mathematical Methods for Physicists*, 6th ed.; Academic Press: San Diego, 2005.

10. Van Siclen, C. Estimating Atomic 1s Orbital Radii. *J. Comput. Appl. Math.* **1987**, *19*, 283–286.

11. Vinet, P.; Smith, J. R.; Ferrante, J.; Rose, J. H. Temperature Effects on the Universal Equation of State of Solids. *Phys. Rev. B: Condens. Matter Mater. Phys.* **1987**, *35*, 1945–1953.